\documentclass[useAMS,usenatbib]{mn2e}

\usepackage{graphicx}
\usepackage{amssymb}

\newcommand{\ion}[2]{\mbox{#1$\,$\textsc{#2}}}

\title[The scattered debris of the Magellanic Stream]{The scattered debris of the Magellanic Stream}
\author[T.~Westmeier and B.~S.~Koribalski]{T.~Westmeier$^{1}$ and B.~S.~Koribalski$^{1}$\\
$^{1}$CSIRO Australia Telescope National Facility, PO Box 76, Epping NSW 1710, Australia}

\begin{document}

  \date{Accepted 1988 December 15. Received 1988 December 14; in original form 1988 October 11}

  \pagerange{\pageref{firstpage}--\pageref{lastpage}} \pubyear{2002}

  \maketitle

  \label{firstpage}

  \begin{abstract}
    Searching the \ion{H}{i} Parkes All-Sky Survey (HIPASS) and its northern extension, we detected a population of very compact high-velocity clouds (HVCs) with similar velocities in the Galactic standard-of-rest frame which appear to be arranged in several filaments aligned with the nearby Magellanic Stream. A comparison with published \ion{O}{vi}/\ion{Ca}{ii} absorption and \ion{H}{i} emission line measurements suggests that the HVCs are condensations within an extended and mainly ionised component of the Magellanic Stream. They coincide in position with a faint gas stream predicted in numerical simulations of the Magellanic Clouds by \citet{Gardiner1996}. Consequently, the Magellanic Stream could be much more extended than generally believed.
  \end{abstract}

  \begin{keywords}
    Magellanic Clouds -- galaxies: interactions -- ISM: clouds.
  \end{keywords}

  \section{Introduction}
  
  Spanning over $100^{\circ}$ across the sky \citep{Putman2003b}, the Magellanic Stream is one of the most prominent \ion{H}{i} structures in the vicinity of the Milky Way. Following several earlier detections (e.g., \citealt{Dieter1965,Wannier1972a}), the Magellanic Stream was first studied over its entire extent by \citet{Mathewson1974} who also noticed its association with the Magellanic Clouds. The northern, leading part of the Magellanic Stream, the so-called Leading Arm, was already known in the early 1970s (e.g., \citealt{Wannier1972b,Mathewson1974}), but its association with the Magellanic Clouds had been hypothetical over many years. Numerical simulations by \citet{Gardiner1996} provided additional evidence for a tidal origin of the Leading Arm in connection with the Magellanic Clouds, recently supported by sensitive \ion{H}{i} observations with the Parkes 64-m telescope \citep{Putman1998,Bruens2005} and metal abundance measurements \citep{Lu1998}. In addition to tidal interaction between the Magellanic Clouds and the Milky Way, ram-pressure may have played an important role in the formation and evolution of the Magellanic Stream (see \citealt{Mastropietro2005} and references therein).
  
  Here we report on the detection of an extended, filamentary population of compact \ion{H}{i} clouds which are likely to be debris associated with the Magellanic Stream. For our analysis we used the \ion{H}{i} Parkes All-Sky Survey (HIPASS; \citealt{Barnes2001}) and its northern extension \citep{Wong2006}. HIPASS was observed with the Parkes 64-m radio telescope and covers the entire sky south of a declination of $\delta = {+2}^{\circ}$. Although being primarily an extragalactic survey, it has been successfully used to study high-velocity gas in the halo of the Milky Way \citep{Putman2002,deHeij2002b}. A fundamental benefit of HIPASS is its relatively high angular resolution of about $15\farcm{}5$ \citep{Barnes2001} which allows for the detection of barely resolved clouds which would be undetectable in other all-sky \ion{H}{i} surveys due to beam dilution. Using the northern extension to HIPASS, we extended the search for high-velocity gas into the northern sky ($\delta < {+25}^{\circ}$) for which only low-resolution data from the Leiden/Argentine/Bonn (LAB) Galactic \ion{H}{i} Survey \citep{Kalberla2005} have been available so far.
  
  \section{Data analysis and results}
  
\begin{figure}
    \centering
    \includegraphics[width=0.8\linewidth]{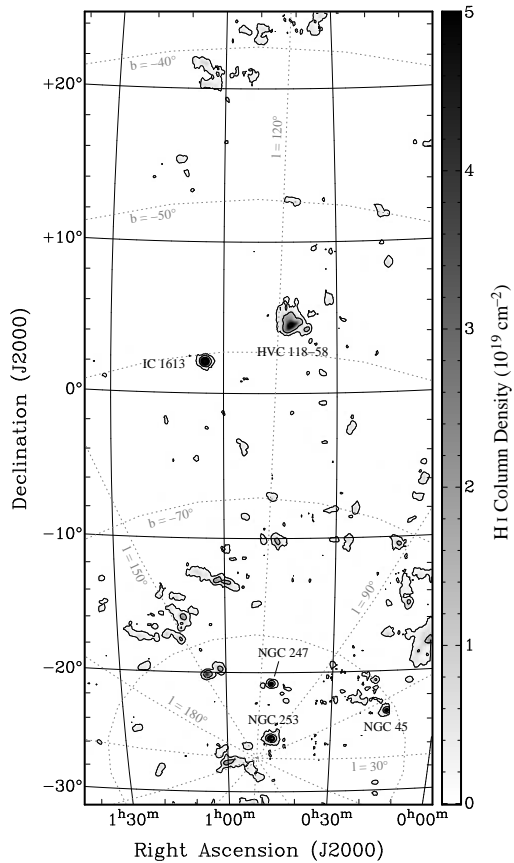}
    \caption{HIPASS \ion{H}{i} column density map of high-velocity gas in the studied region, integrated over the LSR velocity range of ${-415}$ to ${-190}$ and ${+200}$ to ${+495}~\mathrm{km \, s}^{-1}$. The contours are drawn at $10^{18}$, $10^{19}$ and $10^{20}~\mathrm{cm}^{-2}$. Galactic coordinates are overlaid as dotted lines. The only gas at positive velocities are the Sculptor Group galaxies (e.g., NGC~247 and NGC~253 in the southern part of the map). A few areas of the map were affected by artefacts in HIPASS.}
    \label{fig_map}
  \end{figure}

  The original aim of our analysis was to search the HIPASS data cube for a potential \ion{H}{i} filament connecting M31/M33 with the nearby Sculptor Group. The studied region of the sky covers a range of $0^{\rm h}00^{\rm m}$ to $1^{\rm h}30^{\rm m}$ in right ascension and ${-30}^{\circ}$ to ${+25}^{\circ}$ in declination. This corresponds to a total solid angle of $\Omega = 0.36~\mathrm{sr}$, equivalent to almost 1200~square degrees or $3\%$ of the entire sky. The radial velocity range considered in our analysis is $|v_{\rm LSR}| < 500~\mathrm{km \, s}^{-1}$. An \ion{H}{i} column density map of the high-velocity gas in the studied region is shown in Fig.~\ref{fig_map}. We did not find any evidence for a diffuse \ion{H}{i} filament connecting the Local Group with the Sculptor Group within the $3 \sigma$ \ion{H}{i} column density sensitivity of HIPASS of $N_{\rm HI} \simeq 10^{18}~\mathrm{cm}^{-2}$ for warm ($\Delta v = 25~\mathrm{km \, s}^{-1}$) neutral gas emission filling the $15\farcm{}5$ beam. However, the region is very rich in high-velocity clouds (also see \citealt{Putman2002,Putman2003b}). In addition, the Magellanic Stream is running across the lower right corner of the map at low velocities in the local standard-of-rest (LSR) frame. Emission from the Stream can be seen near the right edge of the map at declinations between ${-10}^{\circ}$ and ${-20}^{\circ}$ (also see Fig.~\ref{fig_largemap}). A few galaxies can also be found in the studied region and velocity range, including NGC~247 and NGC~253.
  
  To study the physical properties of the high-velocity clouds (HVCs) in a statistical way we searched the entire data cube by eye for individual objects in the velocity range of $v_{\rm LSR} = {-500} \ldots {-140}$ and ${+60} \ldots {+500}~\mathrm{km \, s}^{-1}$. We compiled a catalogue of 159~objects, six of which could be identified with galaxies in the NASA/IPEC Extragalactic Database (NGC~45, NGC~247, NGC~253, IC~1574, IC~1613 and UGCA~15), leaving us with 153 potential HVCs. Most of the HVCs are very compact with typical angular diameters in the range of only $20'$ to $30'$ FWHM. With their small angular sizes in addition to their isolation and separation from neighbouring clouds some of them are prototypical for the population of compact high-velocity clouds (CHVCs) identified by \citet{Braun1999}. A concentration of HVCs in this area was already noted earlier and partly associated with the nearby Magellanic Stream (e.g., \citealt{Wakker1991}). The HVC catalogue published by \citet{deHeij2002c}, which is based on the Leiden/Dwingeloo Survey \citep{Hartmann1997}, lists only 24~clouds within the region considered in our study, illustrating the importance of high angular resolution for the detection of HVCs and CHVCs. With the HPBW of about $15\farcm{}5$ and the high sensitivity of HIPASS we are able to increase the number of HVC detections in this region by more than a factor six. Almost all of the clouds below the original HIPASS declination limit of $\delta = {+2}^{\circ}$ were previously catalogued by \citet{Putman2002}, but we also found numerous HVCs in the northern part of the survey area, most of which have been unknown before.
  
  \begin{table}
    \centering
    \caption{Mean radial velocities, $\langle v \rangle$, and velocity dispersions, $\sigma$, of the HVC filaments defined in Fig.~\ref{fig_map2} and of the entire population of HVCs identified the in the studied region.}
    \label{tab_fila}
    \begin{tabular}{lrrrr}
      \hline
      Filament & $\langle v_{\rm LSR} \rangle$ & $\sigma_{\rm LSR}$ & $\langle v_{\rm GSR} \rangle$ & $\sigma_{\rm GSR}$ \\
             & [km/s] & [km/s] & [km/s] & [km/s] \\
      \hline
        A & $-322.8$ & $33.3$ & $-197.2$ & $21.3$ \\
        B & $-316.9$ & $52.3$ & $-216.3$ & $51.8$ \\
        C & $-236.8$ & $44.7$ & $-208.8$ & $40.2$ \\
        D & $-229.7$ & $45.0$ & $-176.2$ & $38.7$ \\
        E & $-214.8$ & $35.5$ & $-219.2$ & $36.7$ \\
      \hline
      all & $-261.2$ & $61.4$ & $-203.0$ & $43.5$ \\
      \hline
    \end{tabular}
  \end{table}
  
\begin{figure}
    \centering
    \includegraphics[width=0.88\linewidth]{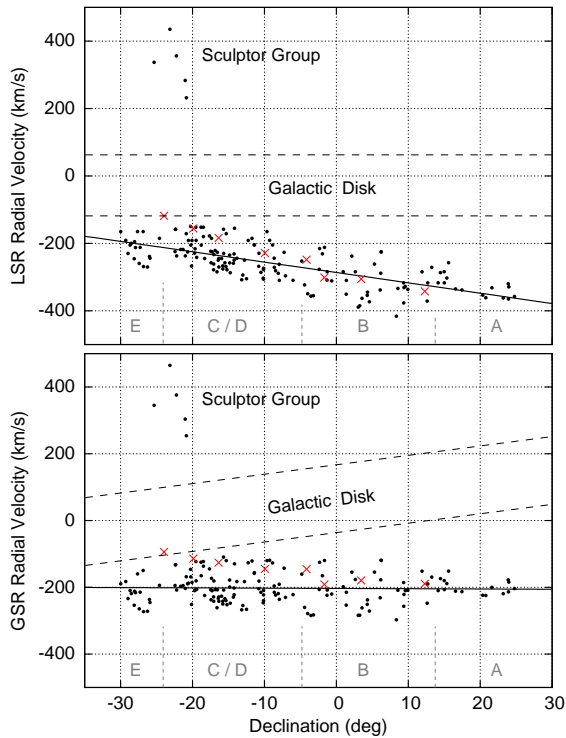}
    \caption{Position-velocity diagrams of the detected HVCs (black dots) in the LSR (top) and GSR (bottom) frames. The five Sculptor Group galaxies found in the studied region at positive velocities are also plotted. The red crosses show the velocities at eight positions along the nearby Magellanic Stream for comparison. The dashed lines enclose the approximate velocity range of \ion{H}{i} emission from the Galactic disc. The solid line shows the result of a linear fit to the velocity distribution of the HVCs. The capital letters indicate the declination ranges of the different groups outlined in Fig.~\ref{fig_map2}.}
    \label{fig_posivelo}
  \end{figure}

  The distribution of radial velocities reveals that all HVCs have negative velocities. We did not find a single HVC with positive velocity in the studied region. In addition, the absolute value of the mean velocity as well as the velocity dispersion decrease when going from the LSR frame to the Galactic standard-of-rest (GSR) frame. The radial velocity distribution of the HVCs as a function of declination is shown in Fig.~\ref{fig_posivelo}. In the LSR frame we observe a strong systematic velocity gradient from north to south. A linear fit to the data points (solid black line) reveals a gradient of ${-3.07} \pm 0.25~\mathrm{km \, s}^{-1}$ per degree. After conversion to the GSR frame this gradient completely disappears, demonstrating that the observed velocity gradient in the LSR frame is completely caused by the rotation velocity of the Galactic disc and that the HVCs have very similar GSR velocities. A linear fit to the GSR data reveals a slope of ${-0.08} \pm 0.25~\mathrm{km \, s}^{-1}$ per degree which is consistent with zero. The mean GSR radial velocity of the HVCs is $\langle v_{\rm GSR} \rangle = {-203.0}~\mathrm{km \, s}^{-1}$ with a very low velocity dispersion of only $\sigma_{\rm GSR} = 43.5~\mathrm{km \, s}^{-1}$. The low dispersion is remarkable given that the detected HVC population is spread over a large area on the sky.
  
  In addition, we did not find a single cloud with $v_{\rm GSR} > {-100}~\mathrm{km \, s}^{-1}$ within the region investigated for this work. All our results suggest that the detected HVCs are part of a homogeneous population of clouds with a common origin which are showing signatures of infall with respect to the Galaxy. No other HVCs of different origin seem to be present in this region although the studied area covers almost 3\% of the entire sky. This result suggests that HVCs and CHVCs could in general be highly clustered instead of being evenly distributed in phase space. In Fig.~\ref{fig_posivelo} we also plotted the velocities at eight positions along the Magellanic Stream as derived from the LAB survey \citep{Kalberla2005}. Apparently, the detected HVCs have velocities very similar to those along the stream in both the LSR and GSR velocity frames, suggesting a connection between the HVCs and the Magellanic Stream.
  
  \begin{figure}
    \centering
    \includegraphics[width=0.544\linewidth]{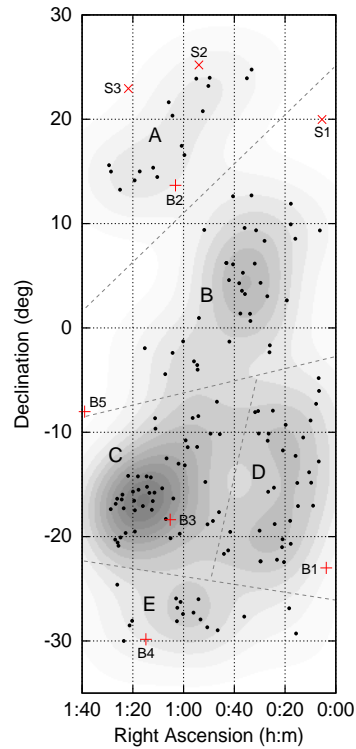}
    \caption{Positions of the individual HVCs (black dots) found in the studied field. The underlying grey-scale image is the result of convolution with a Gaussian of $7\fdg{}5$ FWHM. We identified five major groups labelled with capital letters and separated by dashed lines. The red crosses mark the positions of \ion{O}{vi} absorbers ($\times$) and \ion{Ca}{ii} absorbers ($+$) found by \protect\citet{Sembach2003} and \protect\citet{BenBekhti2008}, respectively (see Table~\ref{tab_abso}).}
    \label{fig_map2}
  \end{figure}
  
  Fig.~\ref{fig_map2} shows a map with the positions of all 153~HVCs identified in the studied field. Obviously, their distribution is not homogeneous. Instead, they form filamentary groups which are separated from each other by gaps of several degrees in width. We separated the HVCs into five groups which are labelled with capital letters in Fig.~\ref{fig_map2}. Most of these filaments have approximately the same orientation angle, running from the north-west to the south-east, providing additional evidence for a common origin of the HVCs. A comparison with the large-scale high-velocity structures in this region (Fig.~\ref{fig_largemap}) shows that the filaments are running almost parallel to the Magellanic Stream, again suggesting that they are related to the stream. The most conspicuous filaments~A and~B remained unnoticed in earlier HVC surveys based on HIPASS which were limited to $\delta < {+2}^{\circ}$.
  
  The kinematic properties of the different filaments are summarised in Table~\ref{tab_fila}, where mean radial velocities, $\langle v \rangle$, and velocity dispersions, $\sigma$, are listed for both the LSR and GSR velocity frames. Although the individual filaments have very similar properties, there are a few subtle differences. Filament~A has a very low velocity dispersion of only $\sigma_{\rm GSR} = 21~\mathrm{km \, s}^{-1}$, whereas the dispersion of filament~B is particularly high with $\sigma_{\rm GSR} = 52~\mathrm{km \, s}^{-1}$. Filament~D is characterised by a significantly lower mean velocity of $\langle v_{\rm GSR} \rangle = {-176}~\mathrm{km \, s}^{-1}$. Filament~C is remarkable for some of its clouds being elongated perpendicular to the orientation of the filament (see Fig.~\ref{fig_map}). This could be explained by ram-pressure stripping under a small angle of attack \citep{Bland-Hawthorn2007}.

  \section{Discussion}
  
  As mentioned before, the positions and velocities of the HVCs as well as the overall structure of the HVC filaments suggest a common origin in connection with the nearby Magellanic Stream. The aggregation of HVCs in this part of the sky was already noted earlier. \citet{Putman2002} identified the overabundance of clouds in this direction as group~1 in their catalogue of CHVCs. In their study of the Magellanic Stream based on HIPASS, \citet{Putman2003b} found HVCs at distances of up to $20^{\circ}$ from the main stream, many of which showed head-tail structures. They suggested that these clouds could represent the clumpiness of the original gas, instabilities along the edges of the stream due to interaction with the Galactic halo, or concentrations within a more extended, mainly ionised stream. For the northern part of their map \citet{Putman2003b} used a preliminary version of the northern HIPASS extension with lower sensitivity (only 20\% of the final integration time) so that many of the compact and faint HVCs in this region and their unique filamentary structure remained unnoticed.
  
  The northern end of filament~A spatially overlaps with the region around M31 and M33 studied by \citet{Braun2004} with the WSRT in total-power mode. They discovered filamentary and clumpy \ion{H}{i} emission with column densities in the range of about $10^{17} \ldots 10^{19}~\mathrm{cm}^{-2}$ over a beam size of $48'$~FWHM. They attributed the gas partly to the Magellanic Stream and partly to a cosmic web filament connecting M31 with M33. The velocity of the gas in the region of our filament~A is in the range of about ${-200} \ldots {-100}~\mathrm{km \, s}^{-1}$ in the Local Group standard-of-rest (LGSR) frame. This is in agreement with the radial velocities of $v_{\rm LGSR} \simeq {-125}~\mathrm{km \, s}^{-1}$ observed for the HVCs at the northern end of filament~A, suggesting that the compact HVCs found in HIPASS could be condensations embedded in a significantly more extended filamentary structure of mainly ionised gas with low neutral column densities associated with the Magellanic Stream.
  
  \begin{figure}
    \centering
    \includegraphics[width=\linewidth]{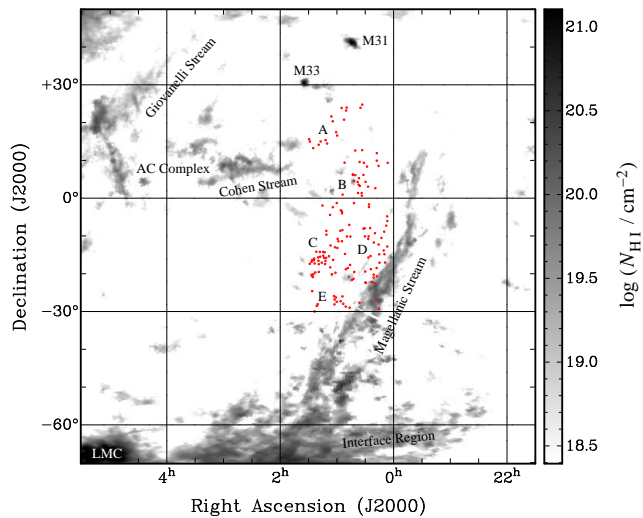}
    \caption{Distribution of the compact HVCs (red dots) in relation to the extended high-velocity structures (grey-scale) in this region of the sky. The underlying map was derived by \protect\citet{Westmeier2007} from the LAB survey \protect\citep{Kalberla2005}. The detected HVC filaments A--E run almost parallel to the Magellanic Stream and in proximity to the Anti-Centre Complex and M31/M33.}
    \label{fig_largemap}
  \end{figure}
  
  In fact, \citet{Sembach2003} detected high-velocity \ion{O}{vi} absorption against three quasars in the northern part of our field (see Fig.~\ref{fig_map2} and Table~\ref{tab_abso} for details). In all three cases the radial velocities of the \ion{O}{vi} absorbers are similar to those of the HVCs found in HIPASS ($v_{\rm GSR} \simeq {-200}~\mathrm{km \, s}^{-1}$), supporting the idea that the HVCs are condensations within a much more extended stream of diffuse gas. In the case of collisional ionisation \ion{O}{vi} has its highest ionisation fraction at a temperature of about $3 \times 10^5~\mathrm{K}$ \citep{Sutherland1993}. Therefore, the \ion{O}{vi} absorbers probably trace the interface region between the warm neutral gas of the Magellanic Stream and the highly ionised Galactic corona. In addition, \citet{BenBekhti2008} discovered intermediate- and high-velocity \ion{Ca}{ii} and \ion{Na}{i} absorption towards several quasars in our field, the velocities of some of which are consistent with those of the HVCs (Fig.~\ref{fig_map2} and Table~\ref{tab_abso}).
  
  A comparison of our observations with the results of numerical simulations obtained by \citet{Gardiner1996} may provide further clues about the origin of the HVC filaments. In their simulations the Magellanic Bridge was formed about 0.2~Ga ago during a very close encounter between the LMC and SMC, whereas the Magellanic Stream and Leading Arm were generated during the previous perigalactic passage of the SMC about 1.5~Ga ago which approximately coincided with another close encounter between the two Magellanic Clouds. Despite being purely gravitational, the simulations of \citet{Gardiner1996} can reproduce the kinematics and morphology of the Magellanic Stream with remarkable precision, although their predictions for the Leading Arm are less accurate. Their Magellanic Stream, however, consists of two separate streams, a denser main stream which can be identified with the one observed in neutral hydrogen, and a secondary stream with much lower surface density which is not observed in \ion{H}{i}. This faint secondary stream runs parallel to the main stream with a separation of about $15^{\circ}$ between the two streams. A comparison with Fig.~6 of \citet{Gardiner1996} reveals that the location of this secondary stream exactly coincides with the population of HVCs found in HIPASS. Therefore, the discovered HVC filaments could be the neutral part of the faint secondary stream predicted by the simulations of \citet{Gardiner1996}. In this case they would have formed about 1.5~Ga ago together with the main filament known as the Magellanic Stream.
  
  \begin{table}
    \centering
    \caption{Summary of \ion{O}{vi} and \ion{Ca}{ii}/\ion{Na}{I} absorbers detected in our field by \protect\citet{Sembach2003} and \protect\citet{BenBekhti2008}, respectively. All lines of sight are plotted in Fig.~\ref{fig_map2}.}
    \label{tab_abso}
    \begin{tabular}{llcrr}
      \hline
        \# & Source           & Ion                      & $v_{\rm LSR}$ & $v_{\rm GSR}$ \\
	   &                  &                          & [km/s]        & [km/s]        \\
      \hline
        S1 & Mrk 335          & \ion{O}{vi}              & $-305$        & $-163$        \\
        S2 & PG 0052$+$251    & \ion{O}{vi}              & $-334$        & $-202$        \\
        S3 & Mrk 357          & \ion{O}{vi}              & $-279$        & $-164$        \\
      \hline
        B1 & QSO J0003$-$2323 & \ion{Ca}{ii}             & $-126$        & $ -96$        \\
	   &                  & \ion{Ca}{ii}             & $-112$        & $ -82$        \\
	   &                  & \ion{Ca}{ii}             & $ -98$        & $ -68$        \\
        B2 & QSO J0103$+$1316 & \ion{Ca}{ii}             & $-351$        & $-248$        \\
        B3 & QSO J0105$-$1846 & \ion{Ca}{ii}/\ion{Na}{I} & $+108$        & $+126$        \\
	   &                  & \ion{Ca}{ii}             & $+167$        & $+185$        \\
	   &                  & \ion{Ca}{ii}             & $+192$        & $+210$        \\
        B4 & QSO B0112$-$30   & \ion{Ca}{ii}/\ion{Na}{I} & $ -13$        & $ -32$        \\
        B5 & QSO J0139$-$0824 & \ion{Ca}{ii}             & $-100$        & $ -70$        \\
      \hline
    \end{tabular}
  \end{table}
  
  Recently, \citet{Yoshizawa2003} carried out more realistic $N$-body simulations of the Magellanic System, including the effects of gas dynamics and star formation. In their best model the secondary stream predicted by \citet{Gardiner1996} is not clearly visible in the gas distribution, although a hint of a very faint and broad gas stream in the corresponding area on the sky may be present in their maps. \citet{Yoshizawa2003} concur with \citet{Gardiner1996} in that the Magellanic Stream was formed during the previous passage of the Magellanic Clouds 1.5~Ga ago. \citet{Mastropietro2005} do not mention the formation of a secondary stream in their simulations, and their map of the expected \ion{H}{i} column density distribution on the sky shows only the main filament of the Magellanic Stream. However, they did not include the SMC in their simulations although it may have played a crucial role in the formation of the Magellanic Stream.
  
  All these simulations are limited by the large errors of previous proper motion measurements for the Magellanic Clouds, introducing uncertainties with respect to the orientation and shape of the orbits of both galaxies about the Milky Way. Recent proper motion measurements for the LMC \citep{Kallivayalil2006a,Pedreros2006} and SMC \citep{Kallivayalil2006b} have reduced the uncertainties significantly. Furthermore, \citet{Besla2007} pointed out that the new proper motion measurements for the LMC and SMC reveal significantly higher space velocities than previously believed, indicating that the Magellanic Clouds might be on their very first passage about the Milky Way. This result was confirmed by \citet{Piatek2008} in their independent reanalysis of the same data. In this astonishing case the evolutionary history of the Magellanic Clouds would have been totally different from what was assumed in all previous simulations, and the Magellanic Stream could not have formed during an earlier perigalactic passage of the Magellanic Clouds.
  
  These uncertainties in the observations and simulations make a more detailed analysis of the origin of the filamentary HVC population near the Magellanic Stream currently impossible. We would like to emphasise the importance of further sensitive observations of the Magellanic System at different wavelengths and with different techniques to provide a comprehensive observational basis for realistic and detailed simulations. This will finally allow us to unravel the origin and evolution of the Magellanic Clouds and their complex system of tidal arms. The gaseous tidal streams are particularly suitable to constrain the evolution of the entire system, and any reliable simulation must succeed in reproducing the details of their complex structure.

  \section{Summary and conclusions}
  
  \begin{itemize}
    \item We detected a population of 153~compact HVCs in HIPASS over an area on the sky of almost 1200~square degrees. They are arranged in several nearly parallel filaments aligned with the nearby Magellanic Stream.
    \item A statistical analysis of their radial velocities reveals that the HVCs have very similar GSR velocities of about ${-200}~\mathrm{km \, s}^{-1}$ with a remarkably small dispersion of only $43.5~\mathrm{km \, s}^{-1}$, suggesting a common origin of the clouds.
    \item A comparison with the filamentary gas found by \citet{Braun2004} suggests that the HVCs could be compact condensations within more extended and mainly ionised gas filaments associated with the Magellanic Stream.
    \item This idea is supported by the detection of \ion{O}{vi} absorption by \citet{Sembach2003} which is thought to trace the interface region between the warm neutral gas of the Magellanic Stream and the ionised Galactic corona.
    \item Consequently, the Magellanic Stream could be much more extended than generally believed. The origin of the extended gas filaments is currently unknown, although a faint gas stream in this region of the sky is predicted by the numerical simulations of \citet{Gardiner1996}.
    \item A more extensive study of HIPASS is under way to map the entire extent of the HVC filaments and investigate their connection with other HVC complexes on the sky.
  \end{itemize}

  \section*{Acknowledgements}

  We thank E.~D.~Skillman and G.~Da~Costa for their inspiration which motivated this project. We also thank L.~Staveley-Smith for helpful comments on the manuscript. The Parkes telescope is part of the Australia Telescope which is funded by the Commonwealth of Australia for operation as a National Facility managed by CSIRO.

  \bibliographystyle{mn2e}
  \bibliography{westmeier.bib}
  
  %\appendix
  
  \bsp
  
  \label{lastpage}
  
\end{document}